\documentclass[aip,longbibliography,pof]{revtex4-1}
\usepackage[bookmarks,colorlinks]{hyperref}

\usepackage{natbib}
\usepackage{bm}
\usepackage{amssymb}
\usepackage{amsbsy}
\usepackage{color}
\usepackage{mathrsfs,mathcomp}
\usepackage{amsmath}
\usepackage{amsfonts}
\usepackage{amssymb}
\usepackage{amsbsy}
\usepackage{color}
\usepackage{mathrsfs}
\usepackage{graphicx}
\usepackage{subfigure}
\usepackage{multirow}
\usepackage{subeqnarray}

\newcommand{\be}{\begin{equation}}
\newcommand{\ee}{\end{equation}}
\newcommand{\bea}{\begin{eqnarray}}
\newcommand{\eea}{\end{eqnarray}}

\begin{document}

\title{Stokes flow in a drop evaporating from a liquid subphase}
\author{Hanneke Gelderblom}
 \email{h.gelderblom@tnw.utwente.nl}
 \affiliation{ 
Physics of Fluids Group, Faculty of Science and Technology, Mesa+ Institute, University of Twente, 7500 AE Enschede, The Netherlands.}
\author{Howard A. Stone}
\affiliation{Department of Mechanical and Aerospace Engineering, Princeton University, Princeton, New Jersey 08544, USA}
\author{Jacco H. Snoeijer}
 \affiliation{ 
Physics of Fluids Group, Faculty of Science and Technology, Mesa+ Institute, University of Twente, 7500 AE Enschede, The Netherlands.}

\date{\today}

\begin{abstract}
The evaporation of a drop from a liquid subphase is investigated. The two liquids are immiscible, and the contact angles between them are given by the Neumann construction. The evaporation of the drop gives rise to flows in both liquids, which are coupled by the continuity of velocity and shear-stress conditions.
We derive self-similar solutions to the velocity fields in both liquids close to the three-phase contact line, where the drop geometry can be approximated by a wedge. We focus on the case where Marangoni stresses are negligible, for which the flow field consists of three contributions: flow driven by the evaporative flux from the drop surface, flow induced by the receding motion of the contact line, and an eigenmode flow that satisfies the homogeneous boundary conditions. The eigenmode flow is asymptotically subdominant for all contact angles. The moving contact-line flow dominates when the angle between the liquid drop and the horizontal surface of the liquid subphase is smaller than $90^\circ$, while the evaporative-flux driven flow dominates for larger angles. A parametric study is performed to show how the velocity fields in the two liquids depend on the contact angles between the liquids and their viscosity ratio. 
\end{abstract}

\maketitle
\section{Introduction}\label{introduction}
The evaporation of a liquid drop from a flat solid substrate has been studied extensively, for example in the context of evaporation-driven particle deposition \cite{Deegan:1997,Deegan:2000, Marin:2011, Yunker:2011} and evaporative cooling \cite{Ristenpart:2007, Dunn:2009, Sefiane:2011}.  However, apparently little is known about the evaporation of liquid drops from a liquid subphase. On a liquid subphase the contact line of the drop is not pinned, the subphase is deformable, and the evaporation can generate a flow in both liquid phases. 
Examples of such a system are gasoline or oil drops on water, or water drops on mercury. Similar interface deformations occur for drops evaporating from soft, deformable solid substrates \cite{Style:2012p3786, Marchand:2012, Lopes:2012, Style:2012}, for which the flow geometry is reminiscent of drops floating on another liquid subphase. 

Evaporation of sufficiently small sessile drops under atmospheric conditions is usually governed by the diffusive transport of vapour in the surrounding air \cite{Deegan:1997, Hu:2002, Popov:2005, Eggers:2010, Cazabat:2010, Gelderblom:2011, Sobac:2011}. The evaporative flux from the drop surface drives a radially outward flow inside the drop, which is responsible for the so-called coffee-stain effect  \cite{Deegan:1997, Deegan:2000}. In drops where the contact line is free to move, there is an additional contribution to the flow field that comes from the receding motion of the contact line \cite{Berteloot:2008, Eggers:2010}. Furthermore, the non-uniform evaporation of sessile drops leads to temperature gradients on the liquid-air interface. The subsequent Marangoni flows can give rise to additional circulation \cite{Ristenpart:2007, Hu:2005p3972, Hu:2006}. 

It has remained a challenge to characterize the flow near the contact line, due to the small length scales involved and the evaporative singularity. Recently, we have studied the nature of the flow near the pinned contact line of a liquid drop evaporating from a flat solid surface by approximating the drop profile by a wedge geometry \cite{Gelderblom:2012}. We have shown that the Stokes flow in this wedge-shaped region can be described by similarity solutions and consists of three contributions: flow driven by the evaporative flux from the drop surface, flow induced by the downward motion of the liquid-air interface, and an eigenmode flow that satisfies the homogeneous boundary conditions and can give rise to Moffatt corner eddies \cite{Moffatt:1964}.
 
\begin{figure}
	\includegraphics[width=15 cm]{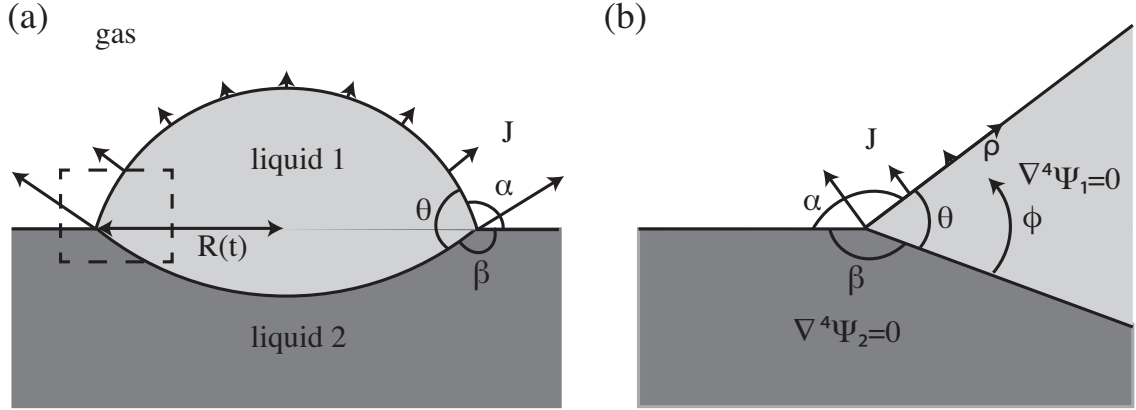}
\caption{(a) A drop of liquid 1 evaporating from a non-volatile liquid subphase (liquid 2). The angles $\theta$, $\alpha$, and $\beta$ between the two liquids can be found from the Neumann construction at the three-phase contact line. The dashed square marks the area close to the contact line, where the drop shape can be approximated by a wedge. (b) An overview of the wedge geometry containing both liquid 1 and liquid 2. The contact line is located at the origin of the polar coordinate system ($\rho$, $\phi$). The evaporative flux $J$ (indicated by the arrows) from the surface of liquid 1 drives the flow inside liquid 1, which is coupled to the flow in liquid 2 via the continuity of velocity and shear stress conditions.}\label{fig1}
\end{figure}

Here, we investigate the evaporation of a liquid drop from a liquid subphase, as shown in Fig. \ref{fig1}(a). We consider drops with a size smaller than the capillary length. In that case, the interface of the liquid subphase remains horizontal and the drop attains the shape of a lens. The angles $\theta$, $\alpha$, and $\beta$ between the liquids are dictated by the three surface tensions in the system via the Neumann construction, such that both the horizontal and vertical components of the capillary forces add to zero \cite{deGennes}. 
We focus again on the nature of the flow near the contact line of the drop, where the drop shape can be approximated by a wedge geometry, see Fig. \ref{fig1}(b).  The evaporation of liquid 1 will cause a flow in both liquids 1 and 2. We describe these flows by deriving self-similar solutions to the Stokes equations in a wedge geometry, in which the flows in both phases are coupled dynamically. Again, the flow in the wedge consists of three contributions: the evaporative-flux driven flow, the flow due to the receding motion of the contact line, and the eigenmode flow for the coupled phases. The eigenmode flow in a similar double-wedge region was already investigated by Anderson \& Davis \cite{Anderson:1993}. They found conditions under which Moffatt eddies can be present in viscous flow of two fluids near a corner of two rigid planes. The present work is a variation of this problem, where there are two free surfaces and one of the liquids evaporates. A parametric study is performed to investigate how much flow is generated in the liquid subsphase by the evaporation of the liquid drop as a function of the angles $\theta$, $\alpha$, and $\beta$, and the viscosity ratio of the two liquids.
We first treat the three flow contributions separately, and then identify the regime in which each contribution is asymptotically dominant. 

\section{Problem formulation}
The geometry of the drop close to the contact line can be approximated by  a two-dimensional wedge, see Fig. \ref{fig1}(b). The three contact angles between the liquids in the wedge add to $\theta+\alpha+\beta=2\pi$.
To ensure both horizontal and vertical balances between the surface tensions at the three-phase contact line, these angles are restricted to $\theta,\alpha, \beta<\pi$ .
We adopt a polar coordinate system ($\rho,\phi$) with the origin located at the contact line and define the velocities in terms of streamfunctions $\Psi_1(\rho,\phi)$ and $\Psi_2(\rho,\phi)$ as 
\begin{eqnarray}
{u_\rho}_1(\rho,\phi)=-\frac{1}{\rho}\frac{\partial \Psi_1}{\partial \phi}, ~{u_\phi}_1(\rho,\phi)=\frac{\partial \Psi_1}{\partial \rho},~{u_\rho}_2(\rho,\phi)=-\frac{1}{\rho}\frac{\partial \Psi_2}{\partial \phi}, ~{u_\phi}_2(\rho,\phi)=\frac{\partial \Psi_2}{\partial \rho}. \label{upsi}
\end{eqnarray}
The flow is governed by the Stokes equations, or equivalently, in terms of the streamfunctions, by the biharmonic equations 
\begin{equation}
\nabla^4\Psi_1=0, \quad \nabla^4\Psi_2=0. \label{Stokes}
\end{equation}
The general self-similar solutions to the biharmonic equations in the two liquids are \cite{Michell:1899}
\begin{subeqnarray}\label{sol}
\Psi_1(\rho,\phi)&=&\rho^m\left[a_1 \cos m\phi+a_2 \sin m \phi +a_3 \cos(m-2)\phi+a_4\sin(m-2)\phi\right]\slabel{sol1},\\
\Psi_2(\rho,\phi)&=&\rho^n\left[c_1 \cos n\phi+c_2 \sin n \phi +c_3 \cos(n-2)\phi+c_4\sin(n-2)\phi\right]\slabel{sol2},
\end{subeqnarray}
where $m,n \neq 0,1,2$; in that case, the form (\ref{sol}) is degenerate and additional terms arise.
The exponents $m$ and $n$ will be selected by the boundary conditions. 

The solution to the flow field described by (\ref{Stokes}) consists of three contributions \cite{Gelderblom:2012}. One contribution comes from the evaporative mass flux from the free surface of liquid 1, which drives a flow inside the drop. A second contribution arises from the volume change of the drop due to the evaporative mass loss. In case the contact line of the drop is pinned, there will be a corresponding contribution to the flow field due to the downward motion of the liquid-air interface \cite{Gelderblom:2012}. On a liquid subphase however, the contact line of the drop is not pinned. Instead, there is a receding motion of the contact line, which results in a decrease in the drop radius and gives a contribution to the flow field. The third contribution to the flow field is the classical  ``eigenmode'' solution to the homogeneous problem for which all boundary conditions are zero, as described by Dean \& Montagnon \cite{Dean:1949}, Moffatt \cite{Moffatt:1964}, and Anderson \& Davis \cite{Anderson:1993}.
From now on, we will refer to these contributions as the \emph{flux solution}, the \emph{moving contact-line solution}, and the \emph{eigenmode solution}. Since (\ref{Stokes}) is a linear system, we can consider each of these three contributions separately. The full solution can be obtained by superposition. 

\subsection{Flux solution}
For the evaporative-flux driven flow, we impose as boundary conditions that
liquids 1 and 2 are immiscible and the geometry is fixed. At the interface between liquids 1 and 2, located at $\phi=0$ (or $\phi=2\pi$) this gives
\begin{subeqnarray}\label{bc2}
&u_{\phi_1}(\rho,0)&=0,\slabel{bc1.2}\\
&u_{\phi_2}(\rho,2\pi)&=0.\slabel{bc2.2}
\end{subeqnarray}
The liquid-air interfaces, located at $\phi=\theta$ (liquid 1) and $\phi=\theta+\alpha$ (liquid 2), are shear-stress free
\begin{subeqnarray}\label{nostress}
&\tau_{\rho\phi_1}(\rho,\theta)&=0,\\ 
&\tau_{\rho\phi_2}(\rho,\theta+\alpha)&=0.
\end{subeqnarray}
The coupling between liquids 1 and 2, at the boundary $\phi=0$, is made by the continuity of velocity and shear stress conditions at the liquid-liquid interface,
\begin{subeqnarray}\label{coup}
&u_{\rho_1}(\rho,0)&=u_{\rho_2}(\rho,2\pi),\slabel{coup1}\\
&\tau_{\rho\phi_1}(\rho,0)&=\tau_{\rho\phi_2}(\rho,2\pi),\nonumber\\
\mathrm{or} &\eta_1\left[\rho \frac{\partial}{\partial \rho}\left(\frac{u_{\phi_1}}{\rho}\right)+\frac{1}{\rho}\frac{\partial u_{\rho_1}}{\partial \phi}\right]_{\phi=0}&=\eta_2\left[\rho \frac{\partial}{\partial \rho}\left(\frac{u_{\phi_2}}{\rho}\right)+\frac{1}{\rho}\frac{\partial u_{\rho_2}}{\partial \phi}\right]_{\phi=2\pi},\slabel{coup2}
\end{subeqnarray}
with $\eta_1$ and $\eta_2$ the dynamic viscosities of liquids 1 and 2, respectively. As a consequence, the viscosity ratio $\eta=\eta_1/\eta_2$ will appear as a parameter in the solution. The conditions above are the homogeneous boundary conditions for the coupled system of $\Psi_1$ and $\Psi_2$. The inhomogeneous condition comes from the evaporative flux that drives the flow.
We impose that liquid 1 evaporates, while liquid 2 is assumed non-volatile
\begin{subeqnarray}\label{outflux}
&u_{\phi_1}(\rho,\theta)&=\frac{1}{\rho_\ell}J(\rho),\slabel{evap}\\
&u_{\phi_2}(\rho,\alpha+\theta)&=0,\slabel{nonvol}
\end{subeqnarray}
with $J$ the evaporative flux and $\rho_\ell$ the liquid density. The evaporative flux from the drop surface is known from the solution to the diffusion equation describing the vapor concentration field in a corner geometry \cite{Deegan:1997}
\begin{equation}
\frac{J(\rho)}{\rho_\ell}=A(\alpha)U\rho^{\lambda(\alpha)-1}.\label{evapflux}
\end{equation}
The prefactor $A$ can be obtained from the asymptotics of the full spherical-cap solution (Popov \cite{Popov:2005}), and $\lambda(\alpha)=\pi/2\alpha$.
Here $U=D \Delta c/R\rho_\ell$ is the velocity scale, which is of order $\mu$m/s for water drops under atmospheric conditions \citep{Marin:2011}, 
with $R$ the radius of the drop as indicated in Fig. \ref{fig1}(a), $D$ the diffusion constant for vapor in air, and $\Delta c=c_s-c_\infty$ the vapor concentration difference (in kg/m$^3$) between the drop surface and the surroundings. Hence, for $\alpha>\pi/2$ the evaporative flux diverges as the contact line is approached, while for $\alpha<\pi/2$ the evaporation is suppressed near the contact line.

The eight boundary conditions for the biharmonic equations (\ref{Stokes}) are thus given by (\ref{bc2}-\ref{outflux}).
The inhomogeneous evaporative flux condition (\ref{evap}) is driving the flow in liquid 1, which in turn generates a flow in liquid 2 via the coupling conditions (\ref{coup}). 
From condition (\ref{evap}) we find that exponent in (\ref{sol1}) is $m=\lambda(\alpha)$. The exponent in (\ref{sol2}) is selected by the coupling condition (\ref{coup1}), which implies $n=m=\lambda(\alpha)$. 
The coefficients $a_1$ to $c_4$  in (\ref{sol1}) and (\ref{sol2}) can be found from the boundary conditions (\ref{bc2})-(\ref{outflux}). The system of equations that has to be solved to find these coefficients is
\begin{equation}
\mathsf{C}\mathbf{x}=\mathbf{b},\label{syst}
\end{equation}
with
\begin{equation}
\mathbf{x}=\left(a_1,a_2,a_3,a_4,c_1,c_2,c_3,c_4\right),
\end{equation}
 a vector containing the coefficients $a_1$ to $c_4$. The right-hand side vector $\mathbf{b}$ contains only zeros except for element $b_7=AU$, which comes from the evaporative driving term.
 The coefficient matrix $\mathsf{C}$ is given by
\begin{eqnarray}
\mathsf{C}=\left(
\begin{array}{cccccccc}
 1 & 0 & 1 & 0 & 0 & 0 & 0 & 0 \\
 0 & 0 & 0 & 0 & C_1 & S_1  &C_1 &S_1  \\
 \lambda  C_2 & \lambda  S_2 & (\lambda-2 )C_3 & (\lambda-2 ) S_3 & 0 & 0 & 0 & 0 \\
 0 & 0 & 0 & 0 & \lambda  C_4 & \lambda  S_4  & (\lambda-2 ) C_5& (\lambda-2) S_5 \\
 0 & -\lambda  & 0 & -(\lambda-2)  & -\lambda  S_1  & \lambda  C_1  &- (\lambda -2) S_1 & (\lambda-2 )C_1 \\
 \eta \lambda & 0 &  \eta (\lambda-2) & 0 & -\lambda  C_1  & -\lambda  S_1 & -(\lambda -2) C_1  & -(\lambda-2 ) S_1\\
  \lambda C_2 & \lambda S_2 & \lambda C_3   & \lambda  S_3   & 0 & 0 & 0 & 0 \\
 0 & 0 & 0 & 0 &C_4& S_4 & C_5 & S_5  
\end{array}
\right)\label{detc}
\end{eqnarray}
with 
\begin{eqnarray}
C_1&=& \cos 2 \pi  \lambda,\quad C_2=\cos \lambda\theta, \quad C_3= \cos (\lambda-2)\theta,\nonumber\\
C_4&=&\cos\lambda(\alpha +\theta ),\quad  C_5=\cos (\lambda-2)(\alpha +\theta ), \nonumber\\
S_1&=&\sin 2 \pi  \lambda, \quad S_2=\sin  \lambda\theta, \quad S_3=\sin (\lambda -2)\theta ,\nonumber\\
S_4&=&\sin\lambda(\alpha +\theta ),\quad S_5=\sin (\lambda-2)(\alpha +\theta ).
\end{eqnarray}
A solution to (\ref{syst}) can be found when det$(\mathsf{C})\neq 0$. Critical points in the solution to (\ref{syst}) appear for those values of $\alpha$, $\theta$, $\eta$ where det$(\mathsf{C})=0$.  In these cases the eight boundary conditions are no longer linearly independent and the eigenmode solution comes into play \cite{Gelderblom:2012}, as will be discussed in Section \ref{seceig}. 

\begin{figure}
	\includegraphics[width=13 cm]{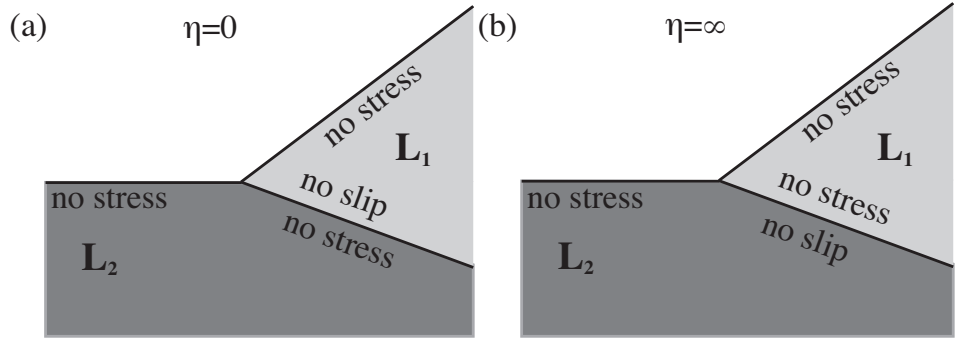}
\caption{Schematic of the boundary conditions experienced by liquids 1 and 2 in the two limiting cases. (a) $\eta=0$: Liquid 1 experiences a no-stress condition at the free surface and a no-slip condition at the interface with liquid 2, whereas liquid 2 experiences two no-stress conditions. (b) $\eta\to\infty$: Liquid 1 experiences two no-stress conditions whereas liquid 2 obeys a no-stress condition at the free surface and a no-slip condition at the interface with liquid 1.}\label{conditions}
\end{figure}

The $8\times8$ matrix given by (\ref{detc}) describes the coupled flow in the two liquids. We anticipate that a decoupling into two $4\times4$ systems arises for $\eta=0$ and for $\eta\to\infty$. In these limiting cases, the flows in the two liquids are completely independent. The boundary conditions experienced by each of the two liquids in these decoupled cases are sketched in Fig. \ref{conditions}. For $\eta=0$, liquid 1 effectively experiences a no-slip condition at the interface with liquid 2, as if it were contacting a solid substrate. This feature is due to the extremely high viscosity of the liquid subphase. For $\eta\to\infty$, the stresses in liquid 2 are so small that effectively a no-stress condition applies and the interface is a free surface. For liquid 2 it is exactly the other way around: no-stress at the interface with liquid 1 when $\eta=0$ and no-slip when $\eta\to\infty$.

\subsection{Moving contact-line solution}
As the drop evaporates its volume decreases and the contact line recedes. The resulting translation of the interface of liquid 1 gives rise to kinematic boundary conditions at the interface between liquids 1 and 2 (located at $\phi=0$) and the free surface of liquid 1 (located at $\phi=\theta$)
\begin{subeqnarray}\label{mov}
u_{\phi_1}(\rho,0)&=&\frac{dR}{dt}\sin (\alpha+\theta)=u_{\phi_2}(\rho, 2\pi),\slabel{mov1}\\
u_{\phi_1}(\rho,\theta)&=&\frac{dR}{dt}\sin \alpha,\slabel{mov2}
\end{subeqnarray}
with $R(t)$ the radius of the drop as indicated in Fig. \ref{fig1}(a), and $dR/dt$ the contact-line speed.
We solve the biharmonic equations describing the flows in both liquids subject to the boundary conditions (\ref{mov}) and (\ref{nostress}-\ref{coup}), (\ref{nonvol}). Since (\ref{mov1}) and (\ref{mov2}) are independent of $\rho$, we obtain for the exponents in (\ref{sol}) $m=n=1$. For $m,n=1$, the solutions (\ref{sol1}) and (\ref{sol2}) are degenerate, and the solutions to the biharmonic equations are of the form \cite{Moffatt:1964}
\begin{subeqnarray}\label{sold}
\Psi_1(\rho,\phi)&=&\rho\left[a_1 \cos \phi+a_2 \sin \phi +a_3 \phi\cos \phi+a_4\phi \sin\phi\right]\slabel{sold1},\\
\Psi_2(\rho,\phi)&=&\rho\left[c_1 \cos \phi+c_2 \sin  \phi +c_3 \phi\cos\phi+c_4\phi \sin\phi\right]\slabel{sold2}.
\end{subeqnarray}
By solving for the coefficients in (\ref{sold1}) and (\ref{sold2}) using boundary conditions (\ref{mov}) and  (\ref{nostress}-\ref{coup}), (\ref{nonvol}), we obtain
\begin{eqnarray}
\Psi_1(\rho,\phi)=\Psi_2(\rho,\phi)=\frac{dR}{dt} \rho  \sin(\alpha +\theta -\phi)\label{sold1e}.
\end{eqnarray}
This result is simply a uniform flow away from the contact line. 

To find an expression for the contact-line speed $dR/dt$ we can again use the solution to the evaporative outer problem \cite{Popov:2005}
\begin{equation}
\frac{dV}{dt}=-\pi R \frac{ D \Delta c}{\rho_\ell} f(\alpha),\label{dvdt}
\end{equation}
with $V$ the drop volume. An expression for $f$ can be obtained from the integrated evaporative flux from a spherical cap and is given by\cite{Gelderblom:2011}
\begin{equation}
f(\alpha)=
\frac{\sin (\pi-\alpha)}{1+\cos (\pi-\alpha)}+4\int_0^\infty \frac{1+\cosh 2(\pi-\alpha)\tau}{\sinh 2\pi\tau}\tanh\left(\alpha\tau\right)d\tau.
\end{equation}
In contrast to the model of Popov \cite{Popov:2005}, the drop geometry on a liquid subphase consists of two spherical caps instead of one. 
In this case, the drop volume is given by
\begin{equation}
V(\alpha,\theta)=\pi R^3\left[\frac{-\cos^3(\alpha)+3\cos(\alpha)+2}{3\sin^3\alpha}+\frac{\cos^3(\alpha+\theta)-3\cos(\alpha+\theta)-2}{3\sin^3(\alpha+\theta)}\right]=\pi R^3h(\alpha,\theta).\label{vol}
\end{equation}
Assuming that contact angles $\alpha$ and $\theta$ remain constant as the contact line recedes, we obtain an equation for the rate of change of the drop radius
\begin{equation}
\frac{dR}{dt}=-\frac{1}{3}U\frac{f(\alpha)}{h(\alpha,\theta)},\label{drdt}
\end{equation} 
where $U=D \Delta c/R\rho_\ell$ is again the characteristic velocity, which is known for a given drop geometry. From (\ref{drdt}) we obtain a prediction for the evolution of the drop radius in time that could be checked experimentally
\begin{equation}
R(t)=\sqrt{\frac{2D\Delta c f(\alpha)}{3\rho_\ell h(\alpha,\theta)}\left(t_f-t\right)},\label{rtime}
\end{equation}
with $t_f$ the total evaporation time of the drop.

\subsection{Eigenmode solution}\label{seceig}
The eigenmode solution is the nontrivial solution to the homogeneous problem \cite{Moffatt:1964}. The form of the solution is still given by (\ref{sol1}) and (\ref{sol2}), but the exponents $m$ and $n$ are selected by the criterion det$(\mathsf{C})=0$, with $\mathsf{C}$ given by (\ref{detc}). In case the determinant of the coefficient matrix $\mathsf{C}$ is equal to zero,  the system of equations is degenerate and one of the equations becomes redundant. In practise, one can find the coefficients 
of the eigenmode flow by setting one coefficient equal to an arbitrary value (here taken unity) and removing one line from the coefficient matrix (\ref{detc}) to remove the redundancy. The last column of the coefficient matrix then serves as the right-hand side vector. This procedure implies that the strength of the eigenmode flow cannot be determined from the ``inner" Stokes flow problem close to the contact line, but is set by the outer flow in the drop \cite{Moffatt:1980, Gelderblom:2012}.

For the standard case of a drop evaporating from a flat solid substrate, the condition for the (Moffatt) eigenmode solution was found to be \cite{Gelderblom:2012}
\begin{equation}
M(\lambda_E,\theta)= \sin 2(\lambda_E-1)\theta-(\lambda_E-1)\sin2\theta=0,\label{standard}
\end{equation}
where $\lambda_E$ is the exponent of the eigenmode solution. To ensure regularity of the velocity field at the origin, only the roots that have $Re(\lambda_E)>1$ are considered. The case $\lambda_E=2$ is a trivial root and does not correspond to an eigenmode. The present case of a drop evaporating from a liquid subphase is more general, since the substrate is a liquid that is not flat. The limit of $\eta$=0 effectively corresponds to a liquid evaporating from a solid substrate, and in this case det$(\mathsf{C})$ reduces to 
\begin{equation}
\mathrm{det}(\mathsf{C})=F(\alpha,\lambda_E)M(\lambda_E,\theta),
\end{equation}
where 
\begin{equation}
F(\alpha,\theta)=-8 \lambda  \sin\left\{\lambda \left[2\pi -(\alpha +\theta )\right] \right\} \sin \left\{\lambda \left[2\pi -(\alpha +\theta ) \right]+2 (\alpha + \theta )\right\}.
\end{equation}
When $\alpha=\pi-\theta$, $F=-8\lambda\sin^2\left(\pi\lambda\right)$ and we exactly recover the case of a liquid evaporating from a flat solid substrate \cite{Gelderblom:2012} given by (\ref{standard}). For arbitrary $\eta$ and $\alpha$ the equation det$(\mathsf{C})=0$ gives the solution to the generalized corner-flow problem of a liquid on a liquid subphase, where, instead of a no-slip condition, a coupling condition between the two liquids applies. Hence, the exponent of the eigenmode solution is a function of the three problem parameters: $m=n=\lambda_E(\alpha,\theta,\eta)$. This problem is a variation of the one studied by Anderson \& Davis \cite{Anderson:1993}, where a two-dimensional viscous flow of two fluids in the corner between two solid walls was studied. Here, instead of two solid walls there are two free surfaces. 

\section{Results}\label{results}
\subsection{Flux solution}
\begin{figure}
	\includegraphics[width=17 cm]{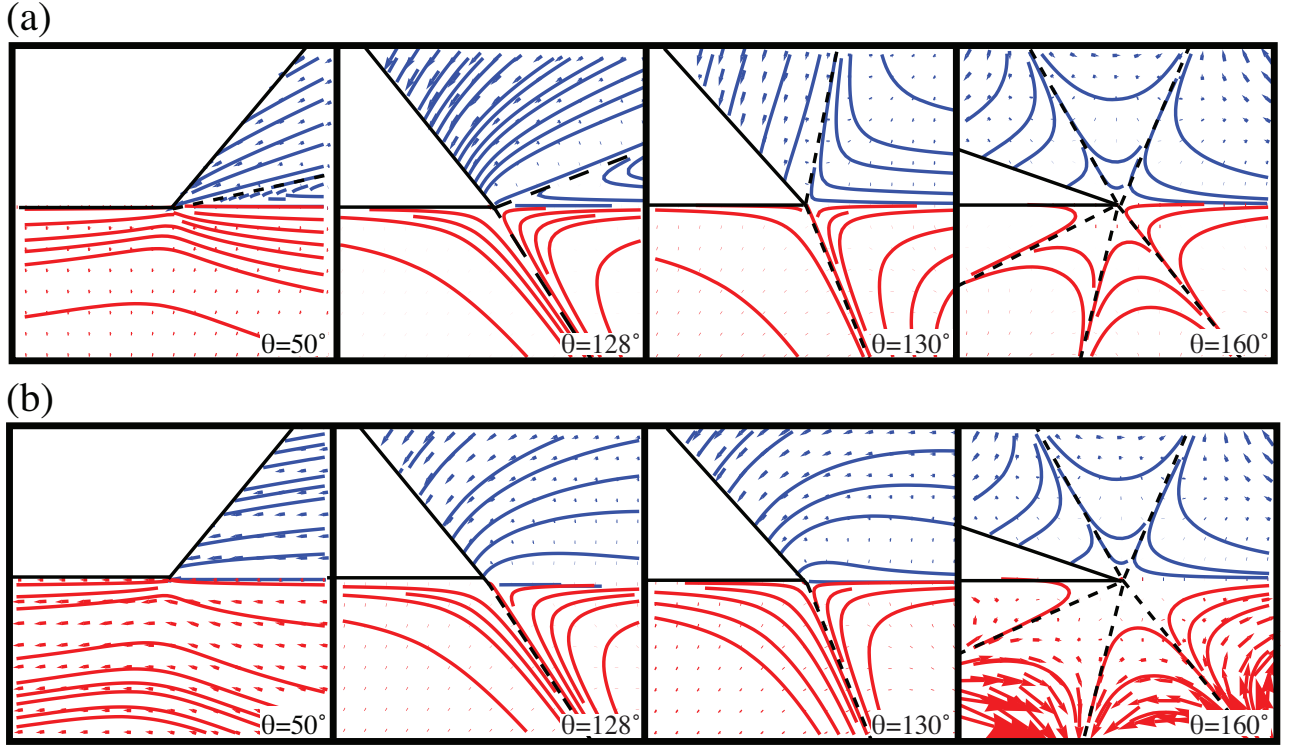}
\caption{(a) Streamline plot of the flux solutions $\Psi_1$ (dark gray, blue online) and $\Psi_2$ (light gray, red online) for a viscosity ratio $\eta=1/10$ (liquid 2 much more viscous than liquid 1). Contact angle $\theta$ is varied from 50$^\circ$ to $160^\circ$, whereas $\alpha$ is adjusted such that the interface of liquid 2 is flat: $\alpha=180^\circ-\theta$. The dashed lines represent the separatrices between the different flow regimes. (b) Same as in (a) but with a viscosity ratio $\eta=10$, which means that liquid 1 is much more viscous than liquid 2. }\label{flatplot}
\end{figure}

The flows in the evaporating drop and the liquid subphase depend on the angles between the phases $\alpha$, $\theta$, and $\beta$, and the viscosity ratio $\eta$. To illustrate the effect of these parameters on the velocity fields in both liquids, we show some representative cases. 
To start we extend our previous work where we studied the flow inside a liquid evaporating from a solid substrate \cite{Gelderblom:2012}.
The current problem is more general, but will tend to the flat solid case when $\eta\ll1$, such that liquid 1 experiences a no-slip condition at the interface with liquid 2, and $\alpha=\pi-\theta$, such that the interface of liquid 2 is flat.
In Fig. \ref{flatplot}(a) we show that in this limit we indeed recover the same flow patterns as described in \cite{Gelderblom:2012}, with the flow in liquid 1 being much stronger than the flow in the more viscous liquid 2. Similarly to the results obtained in our previous work, we observe that for larger contact angles $\theta$ (and hence smaller $\alpha$) the flow shows a sequence of reversal structures. When $\theta=50^\circ$ ($\alpha=130^\circ$), the flow is separated into two regions, such that near the interface with liquid 2, the flow in liquid 1 is directed away from the contact line, towards the center of the drop. The dashed line is the separating flow line that ends in a stagnation point at the contact line. The smaller the angle $\alpha$, the more separatrices appear. Very little flow is generated in the liquid subphase, which is very viscous in the limit $\eta\ll 1$. 
The influence of the liquid subphase becomes apparent when the viscosity ratio $\eta$ is increased, see Fig. \ref{flatplot}(b). When the viscosity of liquid 2 becomes comparable to or smaller than the viscosity of liquid 1, the evaporation of liquid 1 generates a flow in liquid 2 via the coupling conditions (\ref{coup1}) and (\ref{coup2}). The change in viscosity ratio also affects the flow in liquid 1, which now experiences a different boundary condition at the liquid-liquid interface. This results in a disappearance of the separatrices in liquid 1 for $\theta=50^\circ$, $128^\circ$, and $130^\circ$.

\begin{figure}
	\includegraphics[width=13 cm]{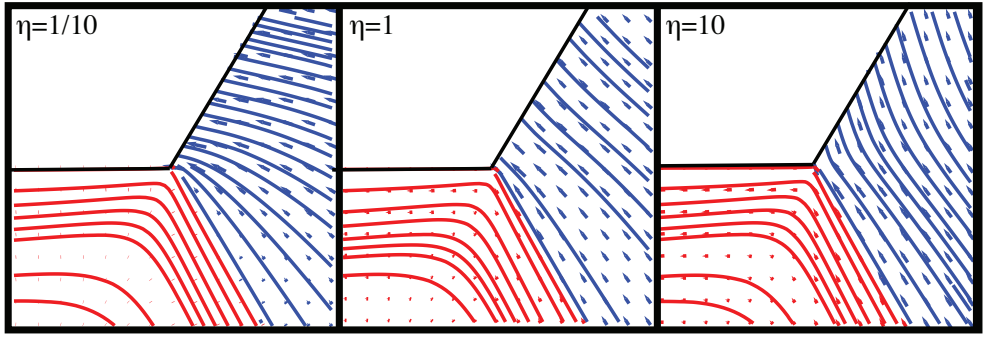}
\caption{Streamline plot of the flux solutions $\Psi_1$ (dark gray, blue online) and $\Psi_2$ (light gray, red online) for $\theta=\alpha=\beta=120^\circ$, which means all surface tensions are equal. The viscosity ratio is varied from $\eta=1/10$ (liquid 1 experiences almost no slip at the interface with liquid 2) to $\eta=10$ (almost free slip).}\label{eta120}
\end{figure}

Note that the cases shown in Fig. \ref{flatplot} are somewhat artificial, since a liquid subphase would not remain flat: the Neumann condition at the contact line will require that $\alpha+\theta > 180^\circ$, to ensure a vertical force balance.
To explore a physically more realistic case, we now consider the situation where all surface tensions are equal, for which $\theta=\alpha=\beta=120^\circ$. In Fig. \ref{eta120} we plot the resulting streamline patterns for three viscosity ratios $\eta$. It can be observed that the velocity field in liquid 1 nearly obeys a no-slip condition for $\eta=1/10$, which means that there is almost no evaporation-driven flow in liquid 2. When the viscosity ratio increases, the evaporation of liquid 1 generates more flow in liquid 2, which can be seen from the larger arrow size in Fig. \ref{eta120} for $\eta=10$ compared to $\eta=1/10$. For $\eta=10$, the flow in liquid 1 follows an almost free-slip condition. The strength of the flow generated in the liquid subphase as a function of the viscosity ratio will be analyzed further below.

\begin{figure}
	\includegraphics[width=17 cm]{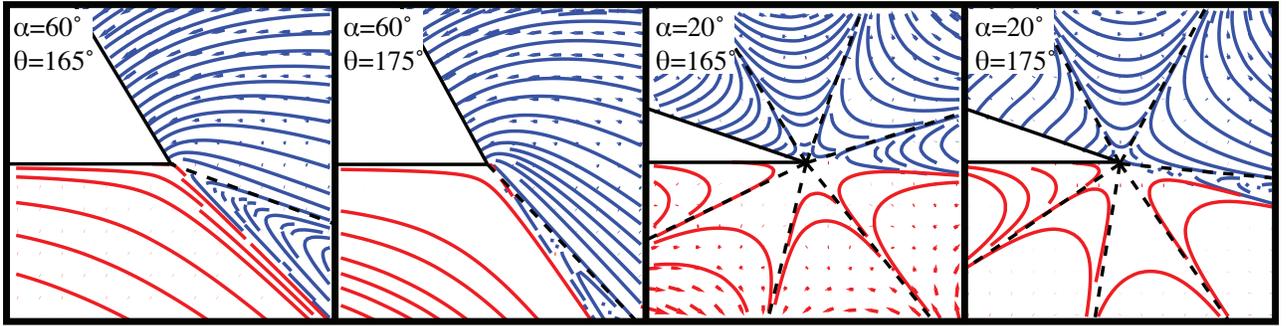}
\caption{Streamline plot of the flux solutions $\Psi_1$ (dark gray, blue online) and $\Psi_2$ (light gray, red online) for $\alpha=60^\circ$ and $\alpha=20^\circ$, where $\theta$ is varied from $165^\circ$ to $175^\circ$, for $\eta=10$. For small $\alpha$, a sequence of wedge-like flow structures is observed in both phases.}\label{exotic}
\end{figure}

In Fig. \ref{eta120} we do not see the reversing flow structures we observed in Fig. \ref{flatplot}. The reason for this feature is that the sequences of wedge-like flow structures shown in Fig. \ref{flatplot} are triggered by large exponents $\lambda$. We can understand this behavior by inspecting the self-similar expressions for the streamfunctions given by (\ref{sol}): since $m=n=\lambda$, a high frequency in the $\phi$-direction arises when $\lambda\gg 1$. Hence,  when $\alpha\ll 90^\circ$, which means $\lambda\gg 1$, multiple wedges appear. To illustrate this behavior we show the flow structures for large and small $\alpha$ in Fig. \ref{exotic}. For $\alpha=60^\circ$, only one separatrix is present in the flow in liquid 1. When $\alpha$ is decreased to $20^\circ$ however, we obtain a sequence of four wedges, which implies that the flow is frequently changing directions.

\begin{figure}
	\includegraphics[width=17 cm]{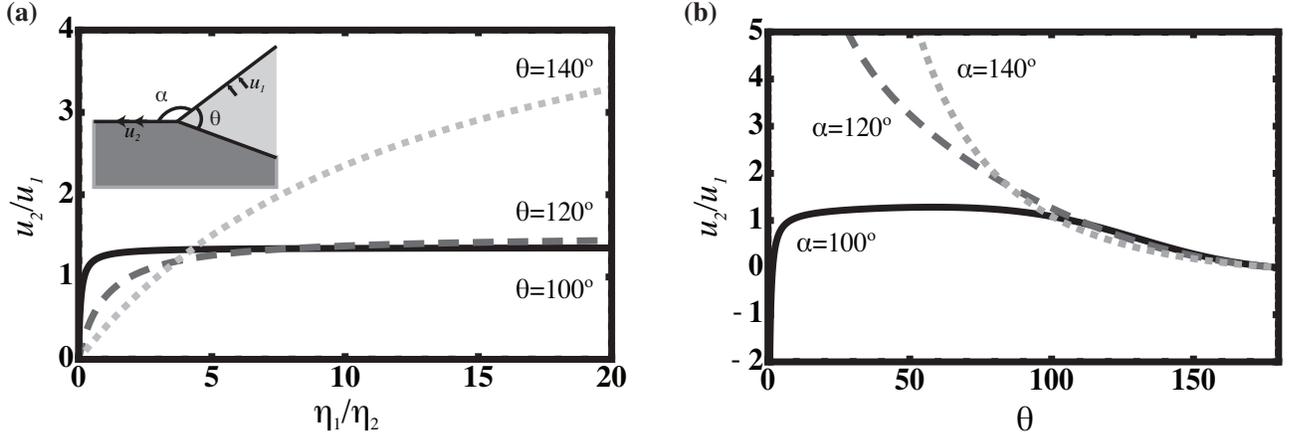}
\caption{(a) Plot of the flow generated at the free surface of liquid 2 compared to the driving by the evaporative flux, $u_2/u_1$, as a function of the viscosity ratio. The inset shows the definition of $u_1 \equiv{u_\phi}_1(\phi=\theta)$, the velocity towards the free surface of liquid 1, and $u_2\equiv {u_\rho}_2(\phi=\alpha+\theta)$, the velocity along the free surface of liquid 2. Here, $\alpha=120^\circ$, whereas $\theta$ is varied from $100^\circ$ (black, solid line) to $120^\circ$ (gray, dashed line) to $140^\circ$ (light gray, dotted line). 
(b) Same velocity ratio as in (a) but now as a function of $\theta$. Parameter $\eta$ is fixed to 1, whereas $\alpha$ is varied from 100$^\circ$ (black, solid line), to 120$^\circ$ (gray, dashed line) to 140$^\circ$ (light gray, dotted line).
}\label{parstud}
\end{figure}

We have seen that the evaporation of liquid 1 generates a flow in liquid 2 through boundary conditions (\ref{coup1}) and (\ref{coup2}). The strength of the flow generated in liquid 2 depends on the viscosity ratio, as shown in Fig. \ref{flatplot} and \ref{eta120}. A measure for the amount of flow generated in liquid 2 by the evaporation of liquid 1 is the ratio ${u_\rho}_2(\phi=\alpha+\theta)/{u_\phi}_1(\phi=\theta)$. This ratio relates the amount of flow generated at the free surface of liquid 2 to the driving term, which is the flow towards the free surface of liquid 1, see the inset in Fig. \ref{parstud}(a). Fig. \ref{parstud} shows how this velocity ratio depends on the parameters $\eta$, $\theta$, and $\alpha$.
In Fig. \ref{parstud}(a) we observe that when $\eta=0$, the flow in liquid 1 obeys a no-slip condition at the interface between the two liquids, which means no flow is generated in liquid 2. In the opposite limit $\eta\to\infty$ a free-slip condition applies, with a maximum amount of flow generated in liquid 2. The  flow strength in liquid 2 depends on the viscosity ratio between the two liquids, and also on the contact angles $\theta$ and $\alpha$. In Fig. \ref{parstud}(b) it is shown that not only the magnitude but also the flow direction depends on $\theta$ and $\alpha$: $u_2/u_1$ can change sign. This sign change is due to the appearance of separatrices in the velocity field, as observed in Fig. \ref{exotic}.

\subsection{Moving contact-line solution}
\begin{figure}
	\includegraphics[width=13 cm]{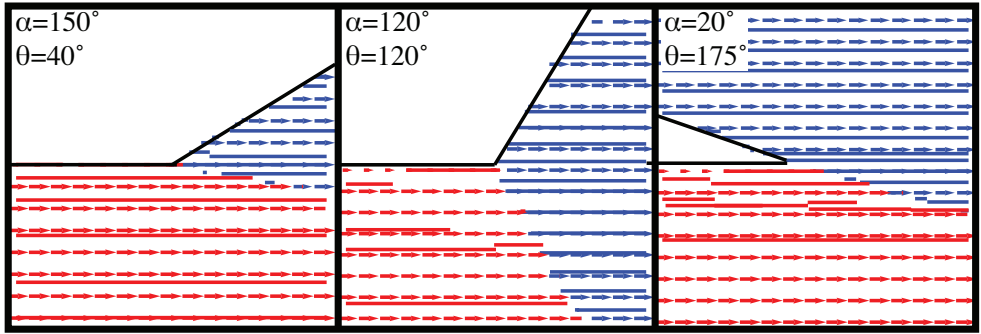}
\caption{Streamline plot of the moving contact-line solutions $\Psi_1$ (dark gray, blue online) and $\Psi_2$ (light gray, red online) for various $\alpha$, $\theta$, and a viscosity ratio $\eta=1$.}\label{uniform}
\end{figure}

The moving contact-line solution is just a uniform flow away from the contact line, as given by (\ref{sold1e}). The corresponding flow pattern is shown in Fig. \ref{uniform}. No gradients in the velocity fields in the two liquids are observed, which means there is no viscous stress. By contrast,  the hinge contribution to the velocity field in a pinned evaporating drop on a solid substrate leads to a significant viscous stress, due to the pinning of the contact line and the no-slip condition at the liquid-solid interface \cite{Gelderblom:2012}.
As the exponents of the streamfunctions for the moving contact-line solution are constant, the flow patterns shown in Fig. \ref{uniform} are the same for all angles $\alpha$ and $\theta$ and viscosity ratios $\eta$. While the exponent of the moving contact-line solution is equal to unity, the exponent $\lambda$ for the flux solution is larger than unity when $\alpha<\pi/2$. Therefore, we anticipate that the moving contact-line solution will be the dominant motion near the contact line when $\alpha<\pi/2$.

\subsection{Eigenmode solution}
\begin{figure}
	\includegraphics[width=17 cm]{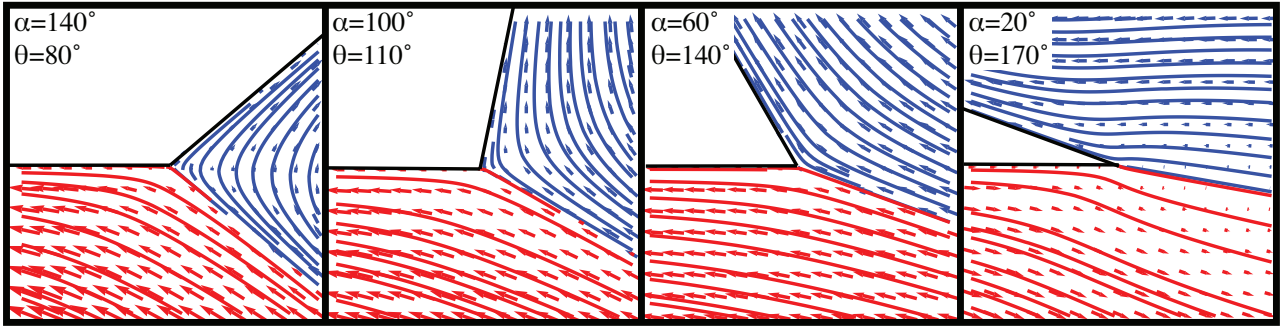}
\caption{Streamline plot of the eigenmode solutions $\Psi_1$ (dark gray, blue online) and $\Psi_2$ (light gray, red online) for decreasing $\alpha$ and increasing $\theta$, for a viscosity ratio $\eta=1$.}\label{eigenmode}
\end{figure}

The streamline patterns of the eigenmode solution are shown in Fig. \ref{eigenmode}. In the plots, we took $\eta=1$, such that the eigenmodes in liquids 1 and 2 are coupled via the boundary conditions (\ref{coup1}) and (\ref{coup2}). We observe that the velocity field in liquid 1 is directed inwards along the interface with liquid 2, and outwards along the free surface. The same flow structure applies to liquid 2, however, for $\alpha=20^\circ$ a reversal in the flow direction in liquid 2 is observed. In that case, the liquid-liquid interface acts as a separatrix. 
For the parameters values used in the plots in Fig. \ref{eigenmode} the exponents of the eigenmode solutions are real-valued, which means no viscous eddies are present \cite{Moffatt:1964}.

\begin{figure}
	\includegraphics[width=17 cm]{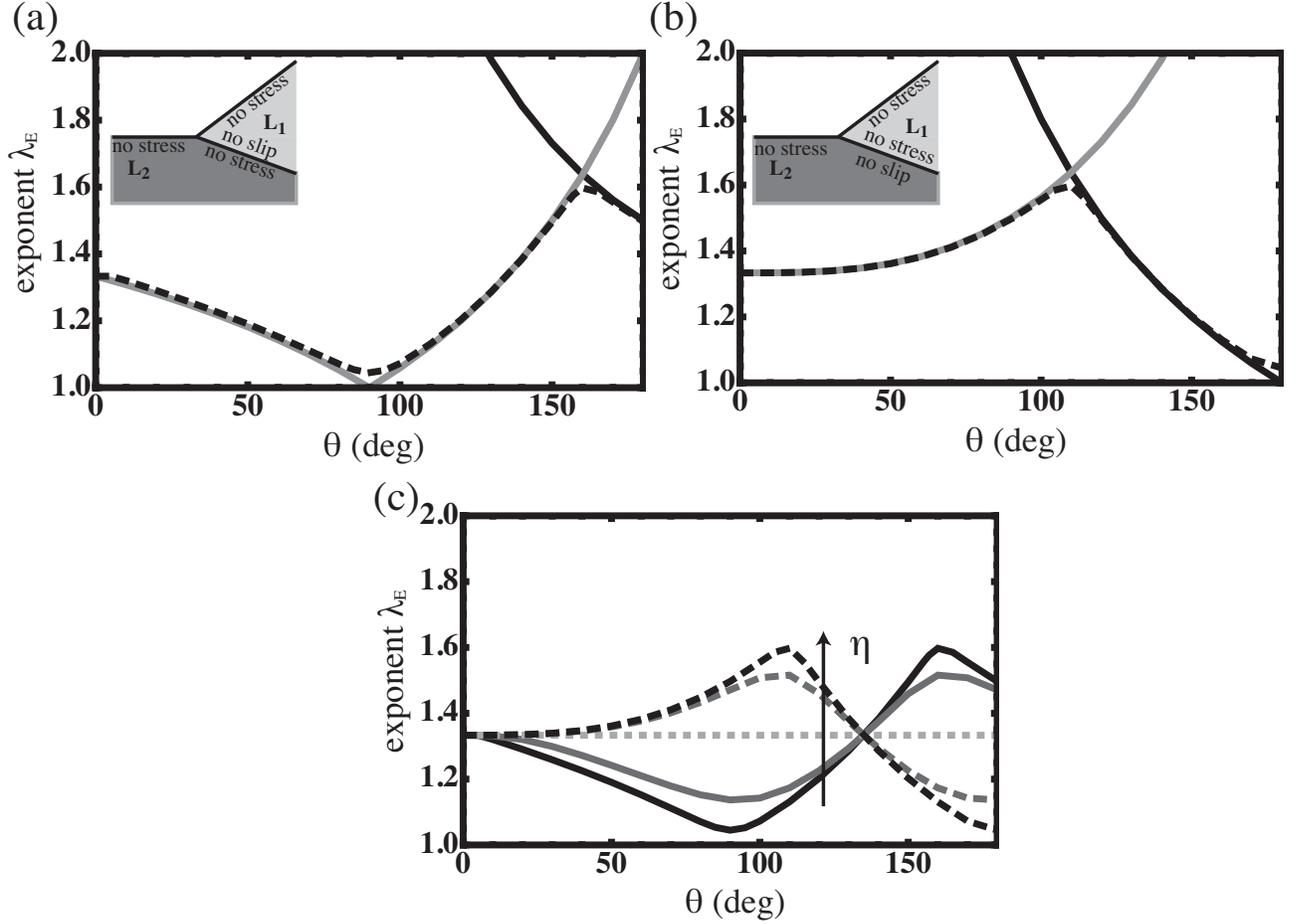}
\caption{(a) Leading-order exponent $\lambda_E$ of the eigenmode solution for liquid 1 (black, solid line) and 2 (gray, solid line) as a function of $\theta$ for $\eta=0$, which means the flow in the two liquids is completely decoupled. The inset shows the boundary conditions experienced by 
liquid 1 and liquid 2 in this case. The black dashed line represents the weakly coupled case where the flow in both liquids obeys the same exponent, for $\eta=1/100$. (b) Leading-order exponent of the eigenmode solutions for the decoupled case where $\eta=\infty$. The black dashed line is the weakly coupled case for $\eta=100$. (c) Leading-order exponent of the coupled eigenmode solution for $\eta=1/100$ (black, solid line), $\eta=1/10$ (gray, solid line), $\eta=1$ (light gray, dotted line), $\eta=10$ (gray, dashed line), and $\eta=100$ (black, dashed line). For all graphs $\alpha=90^\circ$.}\label{exponentseta}
\end{figure}

The exponents $\lambda_E$ of the eigenmode solution follow from the solution to det$(\mathsf{C})$=0 and show multiple branches as a function of $\theta$, $\alpha$, and $\eta$. Asymptotically, the lowest branch that has $Re(\lambda_E)>1$ (to ensure regularity of the velocity field at the origin) dominates. The trivial roots $\lambda_E=1$ and $\lambda_E=2$ have to be excluded, as they do not represent an eigenmode, but are special cases for which the form (\ref{sol}) is degenerate. To understand the structure of the eigenvalues, it is insightful to first describe the limiting cases $\eta=0$ and $\eta\to\infty$ when the eigenmode flows in liquids 1 and 2 are completely decoupled. In these two limiting cases, the leading-order exponents for the flows in both liquids are depicted as a function of $\theta$ in Fig. \ref{exponentseta}(a) and (b) , for a given $\alpha=90^\circ$. The leading-order exponent of the eigenmode flow in liquid 1 is shown as a black solid line and in liquid 2 as a gray solid line. In the decoupled case, the exponents for the eigenmode flows in liquids 1 and 2 are different. The expressions for the exponents are still relatively simple, see for example (\ref{standard}), compared to the case where the flows in liquids 1 and 2 are coupled.

Introducing a weak coupling between the two liquids (dashed lines in Fig. \ref{exponentseta}(a) and (b)) ensures that the flows in both liquids obey the same exponent. This behavior means that the two separate branches represented by the solid lines in Fig. \ref{exponentseta}(a) and (b) give rise to two coupled eigenmodes. At the intersection point where the two separate branches (the solid lines) merge, a double root arises. Consequently, when the flow in the two liquids is coupled, there is a root-splitting behavior of the two coupled eigenmodes at this intersection point. A similar behavior was observed by Anderson \& Davis \cite{Anderson:1993} for a flow of two liquids between two rigid walls. The dominant eigenmode is the one that is closest to the lowest root of the two uncoupled solutions. This leading-order solution is shown as a dashed line in Fig. \ref{exponentseta}. 

To show how the viscosity ratio $\eta$ influences the exponent of the eigenmode flow in the coupled case, we plot the exponent of the eigenmode flow as a function of $\theta$ for different values of $\eta$, for a given $\alpha=90^\circ$ in Fig. \ref{exponentseta}(c). The black solid line represents the leading-order exponent of the coupled eigenmode flow for $\eta=1/100$, which is the same case as depicted in Fig. \ref{exponentseta}(c) as the black dashed line. The opposite case, $\eta=100$, is shown as a black dashed line in both Fig. \ref{exponentseta}(b) and (c). We observe that as the viscosity ratio $\eta$ is increased for $\eta=1/100$ to $\eta=100$, the maximum in the exponent becomes a minimum and vice versa.  

\subsection{Dominant contribution}
\begin{figure}
	\includegraphics[width=9 cm]{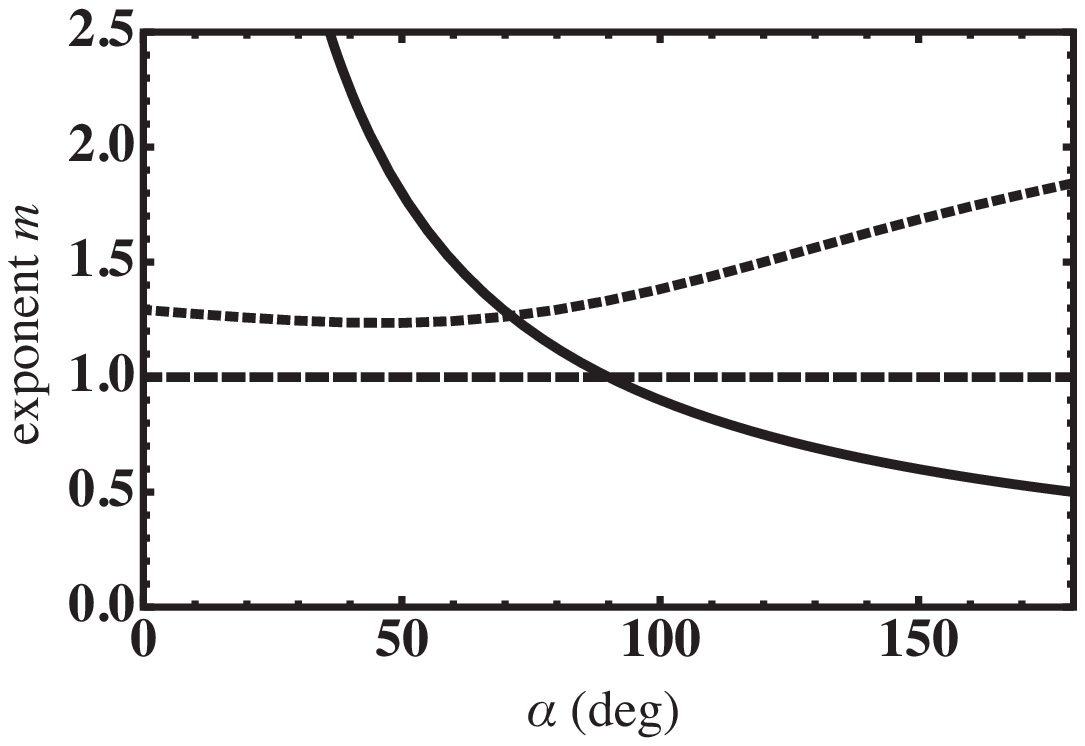}
\caption{Plot of the exponents of the flux (solid line), moving contact line (dashed line), and eigenmode (dotted line) solutions as a function of $\alpha$. For $\alpha<90^\circ$ the moving contact-line solution, for which $m=1$, dominates, for $\alpha>90^\circ$ the flux solution  with $m=\lambda=\pi/2\alpha$ (in radians) is dominant. The leading-order exponent of the eigenmode solution, here plotted for $\theta=120^\circ$ and $\eta=1$, is always subdominant ($m=\lambda_E>1$). }\label{exponent}
\end{figure}

The total evaporation-driven flow in the wedge geometry consists of three contributions: the flux, the moving contact line, and the eigenmode solutions. Each of the three contributions to the total flow in the wedge has a different scaling behavior with distance to the contact line: all are of the type $\rho^m$, but with different $m$. In Fig. \ref{exponent} these exponents are plotted as a function of angle $\alpha$. The moving contact-line solution has exponent $m=1$, which is the smallest of the three exponents for $\alpha<90^\circ$. Therefore, the moving contact-line solution dominates the flow near the contact line for $\alpha<90^\circ$. The exponent for the flux solution (in radians) is $m=\lambda=\pi/2\alpha$. Hence, for $\alpha>90^\circ$, $\lambda<1$ and the flux solution is dominant. The leading-order exponent of the eigenmode solution varies not only with $\alpha$, but also with $\theta$ and $\eta$. Figure \ref{exponent} shows the result for $\theta=120^\circ$ and $\eta=1$. However, for all $\theta$, $\alpha$, $\eta$ the eigenmode solution is always subdominant, since $m=\lambda_E>1$.

\section{Discussion}\label{cpotflow}
We have derived self-similar solutions to the velocity field near the contact line of a liquid drop that evaporates from a liquid subphase.
The nature of the flow strongly depends on the contact angles given by the Neumann construction. In both phases reversing flow structures can be observed, in particular when the angle $\alpha<90^\circ$. We found that there are three contributions to the flow in the wedge geometry: one contribution that is driven by the evaporative flux from the drop surface, one contribution from the receding contact line, and one eigenmode contribution. The amplitudes of the flux and moving contact-line solution are known from the full spherical-cap solution for the vapor transport \cite{Popov:2005}. However, the amplitude of the eigenmode flow can only be obtained by solving the velocity field in the entire drop, which has to be done numerically.
Which of the three flow contributions is asymptotically dominant close to the contact line depends on the contact angle $\alpha$: for $\alpha<90^\circ$ we found that the moving contact-line solution dominates, whereas for $\alpha>90^\circ$, the flux solution is dominant.
The eigenmode solution is always subdominant, in contrast to the drop on a solid substrate (see \cite{Gelderblom:2012}), for which the eigenmode solution is the dominant flow for contact angles larger than $133.4^\circ$.

A fourth contribution to the flow could come from the Marangoni effect: the non-uniform evaporation of the drop leads to temperature gradients on the drop surface, which induces Marangoni stresses \cite{Ristenpart:2007}. These Marangoni stresses can give rise to additional flow circulation in the drop, and also alter the velocity field in the liquid subphase.  In principle, our analysis could be extended to include the Marangoni-driven contribution to the flow in the two liquids following the method of Ristenpart et al. \cite{Ristenpart:2007}

It would be interesting to see if the reversing flow structures found in the present analysis can be measured in an experiment. A direct consequence of our analysis of the moving contact-line solution is the exact expression (including the prefactor) for the time evolution of the drop radius in time (\ref{rtime}), which could easily be verified. To our knowledge, the evaporation time of a drop on a liquid subphase has not been investigated experimentally so far.

We have applied the solutions to the flow in a wedge geometry that contains two liquid phases to the situation of a volatile drop evaporating from a non-volatile liquid subphase. The same solutions also hold for the inverse problem of a non-volatile drop on a volatile subphase, such as oil drops on water. In that case, however, the outer problem for the evaporation is different, since we have a volatile liquid bath instead of a volatile drop. Evaporation will lead to a decrease in the bath level, leaving the drop geometry unchanged. Hence, the prefactor in the expression for the evaporative flux (\ref{evapflux}) will be different. Since there is no contact-line motion in this case, also the eigenmode flow can become important.

\begin{acknowledgements}
H. G. acknowledges the financial support of the NWO-Spinoza program.
\end{acknowledgements}

\end{document}